\documentstyle[psfig]{l-aa} 

\begin{document}

\thesaurus{02        
           (11.03.4 A2390; 
            11.03.3 ; 
	    12.07.1 ; 
	    13.09.1 ; 
	    11.19.5 ; 
	    11.19.6) 
}

\title{
Deep ISOCAM view of the core of the lensing cluster A2390
}

\author{L. L\'emonon\inst{1} \and M. Pierre\inst{1} 
\and C.J. Cesarsky\inst{1}
\and D. Elbaz\inst{1}
\and R. Pell\'o\inst{2}
\and G. Soucail\inst{2}
\and L. Vigroux\inst{1}
}

\offprints{ L. L\'emonon \small ludovic.lemonon@cea.fr}

\institute{DAPNIA/Service d'Astrophysique, CEA/Saclay  F-91191
Gif-sur-Yvette Cedex 
\and Observatoire Midi-Pyr\'en\'ees, Laboratoire d'Astrophysique - 
UMR 5572, 14 Avenue E. Belin, F-31400 Toulouse, France
}

\date{Received ??? , Accepted ???}  

\maketitle

\begin{abstract}

We have imaged the inner square arcminute of the well known lensing and 
cooling flow cluster A2390 
($z = 0.23$) down to a sensitivity of 65 and 130 $\mu$Jy 
at 6.75 and 15 $\mu$m 
, respectively. We report the first evidence of an active star-forming region
 in a 
cooling flow (at those wavelengths) and
 strong emission in the mid-IR from lensed galaxies located at $z=0.9$.

\end{abstract}

\section{Introduction}
The cluster of galaxies A2390 (z=0.228) possesses remarkable properties
which makes its study particularly attractive: presence of a
``straight'' giant gravitational arc (z=0.913), numerous arclets, an
elongated galaxy distribution (\cite{Mellier} ; \cite{Pello}) and a 
large velocity
dispersion (1093 km s$^{-1}$, Carlberg et al., 1996) as well as a high X-ray
luminosity ($\sim1.5 ~10^{45}$ erg s$^{-1}$ in the [0.1--2.4] keV band). A deep
HRI ROSAT pointing revealed an elongated X-ray morphology, the existence
of a secondary maximum responsible for the observed  gravitational shear
in the optical and a strong cooling flow of $\sim800$ M$_{\odot}$yr$^{-1}$
 (Pierre et al., 1996). All this
indicates that A2390, and its underlying gravitational potential, is
especially relevant for our understanding of massive cluster formation,
which is, in a hierarchical scenario, closely related to the history of
galaxy/star formation. This has motivated deep ISOCAM
observations of the cluster core  during the guaranteed time programme
DEEPXSRC. We present here the observations and  results of the cD
galaxy and the $z=0.9$ lensed system.
Throughout the paper we assume $H_{o} = 50$ km s$^{-1}$Mpc$^{-1}$ and $q_{o}$ = 0.5.

\section{ISOCAM observations}
The core of Abell 2390 was observed by ISOCAM, during revolution 393, 
using a 6$\times$6 microscan raster with a step of 
10\arcsec and a pixel size of 3\arcsec. The observations were done in two bands, 
centered on 6.75 $\mu$m (LW2) and 15 $\mu$m (LW3), with an integration time
 of 5 s., including 30 read-outs per raster position. The scan covers
 $2.4' \times 2.4'$ and the maximum sensitivity area (5400 s of 
integration time per pixel position) $46\arcsec \times 46\arcsec$. Thus
images are centered
 on the ``straight arc'', with a resulting pixel size, 
after distortion field correction, of 1\arcsec. Residual 
distortion errors on this corrected map are less than 0.2\arcsec. The Point Spread 
Functions (PSF) FWHM are
about 3.5\arcsec\, in LW2 and 5\arcsec\, in LW3.

The two main steps in the detection of faint sources with
ISOCAM are the removal of cosmic ray glitches and transient 
effects. A wavelet technique method (PRETI, \cite{Starck}) 
has been designed to overcome 
these two effects and was successfully applied to the case of low luminosity
sources. We adapted the method to our data
in the manner described by Aussel et al.$\!$ (1998).
Sources were detected on the final map
correlated with the noise 
map by a wavelet technique, and fluxes calculated by an aperture photometry in a
radius of 3\arcsec.
To achieve a correct calibration of the PRETI photometry algorithm and
to evaluate associated uncertainties, as well as the
completeness of the survey,
 we performed simulations for both channels by adding sources
 ranging from 25 $\mu$Jy to 1 mJy, using a PSF 
model (\cite{Okumura}) and a transient model (\cite{Abergel}).
For these observations, we have a 90\%
completeness level at 65 $\mu$Jy and 130 $\mu$Jy at 6.7 
$\mu$m and 15 $\mu$m, respectively.
The recovered flux is about 35\%at 6.7 
$\mu$m and 20\% at 15 $\mu$m, due 
to the combined effects of detector transients, wavelet reconstruction, 
and PSF.
Relative uncertainties of fluxes decrease from $\sim$50\% 
at 100 $\mu$Jy to $\sim$10\% above 700 $\mu$Jy.

\begin{figure}
\psfig{file=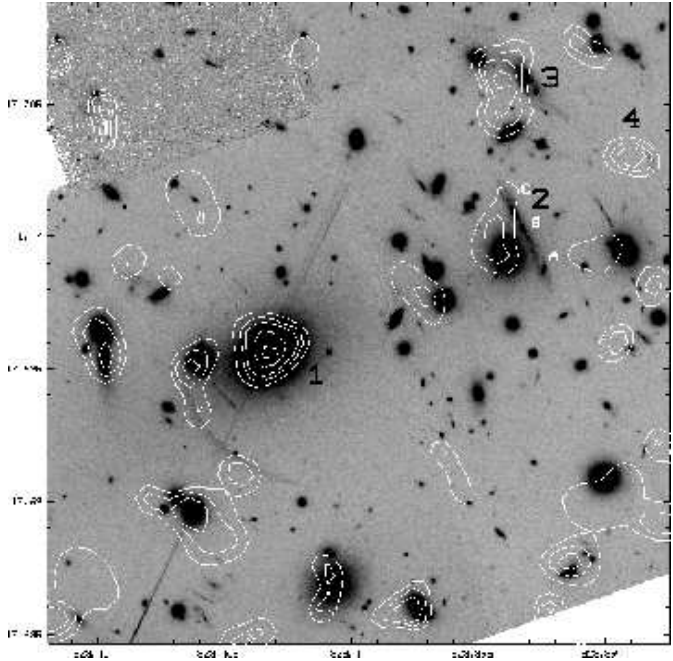,width=8.8cm,angle=0}
\caption []{6.7 $\mu$m image overlaid on the F814W HST view of the core of 
A2390. Wavelet 
contours are drawn at 1, 2, 3, 5 and 7 $\mu$Jy arcsec$^{-2}$
 (with zero-level 
background).
 The straight arc is marginally
detected beside a neighboring cluster-galaxy (\#2). A lensed galaxy (\#3) 
at $z=0.913$ is also detected (``D'' in \cite{Pello}) but the emission is
 clearly contaminated by the 2 
surrounding galaxies. The source \#4 may be associated with a very faint
 optical counterpart.}
\label{lw2}
\end{figure}

\begin{figure}
\psfig{file=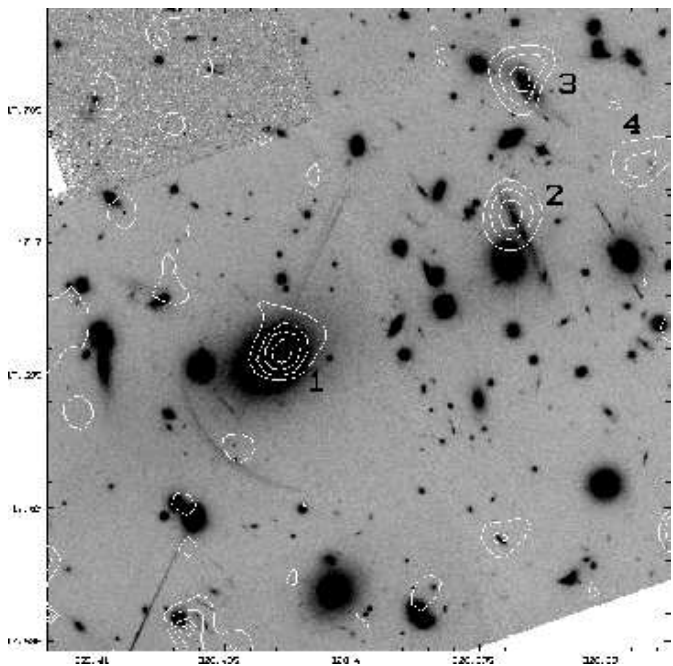,width=8.8cm,angle=0}
\caption []{Same as Fig. \ref{lw2} with 15 $\mu$m contours  at 4, 8, 12 and 16 
$\mu$Jy arcsec$^{-2}$. Here emission from the straight arc is clearly
detected (\#2) but could be contaminated by the neighboring galaxy.}
\label{lw3}
\end{figure}

The 6.7 $\mu$m and 15 $\mu$m rasters were aligned on the HST image of A2390.
 The ISO
 attitude solution is accurate to
within $\sim$6\arcsec, with negligible rotation for this field size, 
and hence mid-IR/optical alignment requires only a translation 
(Aussel, private communication). Superimposing the
cD optical center on the centroid of its nearest 6.7 $\mu$m neighboring
source allowed us to unambiguously identify the brightest sources in
 the ISOCAM field (Fig. \ref{lw2}).
 The same method, applied to the 15 $\mu$m raster, did not yield a 
satisfactory  match 
between ISO and optical sources. This led us to consider the brightest 
non-blended point-like source detected in the maximum sensitivity area of both 
at 6.7 $\mu$m and 15 $\mu$m (source \#4) and impose the 
coincidence of the two centroids. This latter configuration allowed 
 satisfactory optical/IR agreement
(Fig. \ref{lw3}). As a result of the alignment
 process, 6.7 $\mu$m and 15 $\mu$m 
emission peaks do not coincide. The resulting offset between the two centroids
is 2$\pm 1.5$\arcsec\,, approaching the limits on our position accuracy (Fig. \ref{zoom}).
\begin{table*}
\caption[ ]{
6.7 $\mu$m and 15 $\mu$m ISOCAM photometry : uncertainties were estimated from simulations, see text.
 Resulting errors on the alignment
 are less than 0.2\arcsec rms.
* indicates blended
objects.
 V and I magnitudes are from the HST F555W and F814W images while B magnitudes
 are from 
ground-based images (Pell\'o et al., 1998).
}
\label{sources}
\begin{flushleft}
\vspace{-.5cm}
\begin{tabular}{llcccccccccc}
\noalign{\smallskip} \hline \noalign{\smallskip}
ID & name & R.A. & Dec. & 6.7 $\mu$m & 6.7 $\mu$m & 15 $\mu$m & 15 $\mu$m & lensing & V & I & B \\ 
 & & J2000.0 & J2000.0 & adu/g/s & $\mu$Jy & adu/g/s & $\mu$Jy & factor & mag & mag & mag \\
\noalign{\smallskip} \hline \noalign{\smallskip} 
 1 & cD galaxy & 21h53m36.76s & 17$^{\circ}$41\arcmin43.9\arcsec & 0.22 & 300$^{+50}_{-40}$ & 0.22 & 500$^{+80}_{-70}$ & - & 17.5 & 16.2 & 19.4\\
 2 & straight arc & 21h53m34.46s & 17$^{\circ}$42\arcmin03.3\arcsec & $<$0.02 & $<$100* & 0.18 & 440$^{+60}_{-60}$  & 10--20  & 21.5 & 19.8 & 21.9 \\
 3 & object D & 21h53m34.37s & 17$^{\circ}$42\arcmin22.2\arcsec & $<$0.12 & $<$270* & 0.15 & 400$^{+60}_{-50}$ & 2--3 & 22.4 & 19.6 & 23.5 \\ 
 4 & faint object & 21h53m33.16s & 17$^{\circ}$42\arcmin11.0\arcsec & 0.06 & 110$^{+40}_{-60}$ & 0.11 & 350$^{+50}_{-40}$ & - & 25.4 & 23.5 & 26.2 \\ 
\noalign{\smallskip} \hline
\end{tabular}
\end{flushleft}
\end{table*}
\begin{table*}
\caption[ ]{Visible and near-IR SEDs of the 4 relevant ISOCAM sources, compared to a bright E galaxy belonging
to the cluster. Integrated fluxes have been normalized to the HST F814W (I)
. \dag\, indicates photometric redshift.}
\label{SED}
\begin{flushleft}
\vspace{-.5cm}
\begin{tabular}{llclccccccccc}
\noalign{\smallskip} \hline \noalign{\smallskip}
ID & name & z &  & B & g & V & R & r & I$_{F814W}$ & I' & J & K' \\
  & &  & $\lambda_{eff}$ (nm) & 437 & 486 & 545 & 641 & 669 & 799 & 832 & 1237 & 2103 \\ 
\noalign{\smallskip} \hline \noalign{\smallskip} 
 & Cluster galaxy & 0.23 &  & 0.17 & 0.42 & 0.52 & 0.80 & 0.74 & 1.00 & 0.93 & 0.76 & 0.39  \\
 1 & cD galaxy & 0.228 &  & 0.37 & 0.75 & 0.58 & 0.94 & 0.70 & 1.00 & 1.11 & 0.94 & 0.44  \\
     & filament & &  & 0.63 & $<$ 0.2 & 0.97 & 0.29 & 0.18 & 1.00 & $<$ 0.7 & $<$ 0.3 & $<$ 0.1 \\
 2 & straight arc B-C & 0.913 &  & 0.21 & 0.57 & 0.58 & 0.51 & 0.63 & 1.00 & 1.28 & 1.08 & 0.61 \\ 
     & straight arc A & 1.033 &  & 0.45 & 0.76 & 0.15 & 0.72 & 0.47 & 1.00 & 1.18 & 1.06 & 0.38 \\
 3 & object D & 0.913 &  & 0.10 & 0.14 & 0.19 & 0.28 & 0.30 & 1.00 & 0.76 & 1.40 & 0.95 \\
 4 & faint object & $\sim$ 0.4\dag & & 0.31 & - & 0.43 & 0.98 & 0.66 & 1.00 & 0.64 & 1.27 & 2.40 \\
\noalign{\smallskip} \hline
\end{tabular}
\end{flushleft}
\end{table*}
\section{Results and discussion}
In this letter, we restrict the discussion to the 4 sources found in the 
maximum sensitivity area of ISOCAM rasters and seen both at 6.7 $\mu$m and 15 $\mu$m,
 i.e. \#1--4. Their photometric properties are summarized
in Table \ref{sources}. The visible and near-IR spectral energy distributions 
(SED) computed for these 4 sources are given in Table \ref{SED}, and compared 
to those obtained for a typical cluster galaxy. Details on these 
photometric data can 
be found in Pell\'o et al. (1998) and the references therein. In addition,
 about 20 objects are identified
 at 6.7 $\mu$m and 10 at 15 $\mu$m, which will be discussed in a forthcoming paper.
All sources are point-like for ISOCAM, except the cD galaxy
which extends over two times the PSF FWHM at 6.7 $\mu$m (i.e $\sim 20$ kpc), 
and then allow us to exclude a pure AGN emission.

\subsection{The cD galaxy and its cooling flow}
\begin{figure}
\psfig{file=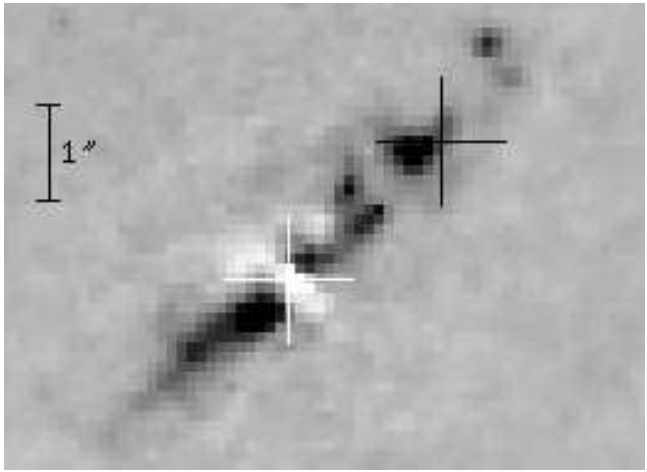,width=8.5cm}
\caption []{The V-I color HST image, obtained from F555W and F814W, of the
 core of the cD galaxy of A2390. 
White and black crosses indicate 6.7 $\mu$m and 15 $\mu$m maximum emission respectively with
the associated uncertainties
(see text).
Some very blue globules suspected to be active star forming regions 
could be responsible for the 15 $\mu$m emission which coincides with the 
most luminous one, and associated with 
 strong cooling flow ($\sim800$ M$_{\odot}$yr$^{-1}$) 
surrounding this cD galaxy.
They are part of a NW--SE filament, aligned with the cD main axis, and 
the overall cluster X-ray elongation on large scale.}
\label{zoom}
\end{figure}

\begin{figure}
\psfig{file=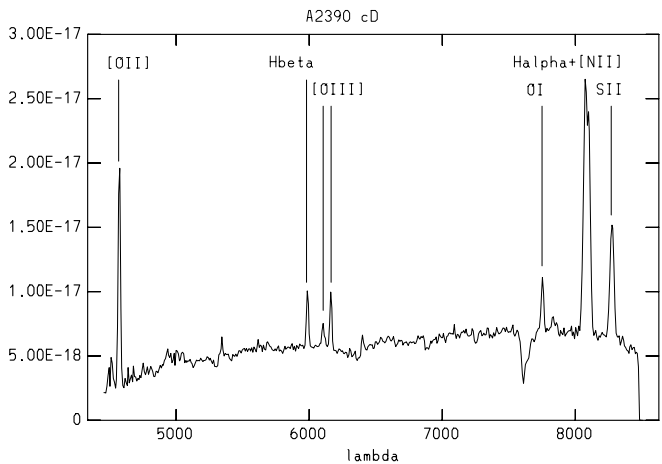,width=9.8cm}
\caption []{Mean long-slit spectrum of the filament within the cD galaxy 
(without reddening correction), from CFHT (\cite{Arnouts}).}
\label{Halpha}
\end{figure}

The cD galaxy is detected both at 6.7 $\mu$m and 15 $\mu$m, with a flux  
of $300^{+50}_{-40} \mu$Jy and $500^{+80}_{-70} \mu$Jy respectively. 
 VLA observations (\cite{Arnouts})
show a point-like source with decreasing radio fluxes of 140, 16 and 
5.5 mJy at 6, 2 and 1.3 cm, respectively. Assuming a power law spectrum, the
 mid-IR flux
 would be some $10^5$ fainter than observed, which excludes a jet-like 
synchrotron contribution to the observed mid-IR emission. 
In galaxies where the mid-IR emission is dominated by an old stellar population, the 
ratio 6.7 $\mu$m/15 $\mu$m is $> 1$.
An excess of 15 $\mu$m emission in field galaxies indicates the presence of 
dust heated 
 by UV photons from star-forming regions.
Compared to the 6.7 $\mu$m/15 $\mu$m ratios observed in other nearby early-type
galaxies (\cite{Suzanne}) or in distant cD galaxies in clusters (\cite{next}),
the ratio of $\sim 0.6$ found here for the cD is exceptional.
 This ratio is compatible
with the colors of the disk component of Centaurus A 
(\cite{Felix}), a nearby giant early-type 
galaxy exhibiting 
active star-forming regions in dust lanes, due to a merge with a spiral
 galaxy. Thus, the cD in A2390 is probably also 
undergoing active star formation.
However, our cD galaxy looks notably different in other wavelengths.
 Cen A shows a jet plus an extended 
radio emission but our cD does not.
In addition, a B image reveals the existence of a filament extended 
along the main axis of the cD, while V and I HST
images show the presence, within the filament, of very blue globules 
possibly associated with the 15 $\mu$m maximum (Fig. \ref{zoom}). 
Strong emission lines are present across the long-slit spectrum of the 
filament 
(Fig. \ref{Halpha})
with ratios indicative of massive
star formation associated with shocks and incompatible with an active nucleus
(\cite{Allen}, \cite{Baldwin}).
Finally the SED of the filament exhibits a clear excess in the V and B bands
 with respect to what is expected for a typical
 elliptical.
 Assuming that the V flux is mainly produced by forming stars, we derive
 $M_V = -20.8 \pm 0.2$ and a 
SFR of $8 \pm 4$ h$_{50}^{-2}$ M$_{\odot}$ yr$^{-1}$ for the optical filament,
 in agreement with the values obtained from the B
flux, corrected for the $[OII]3727$ emission.
No absorption has been considered in this calculation, so 
this estimated SFR has to be taken as a lower limit.
According to the results derived from V and B band, and from IR,
 the optical light is probably 
coming from the most external regions of the star-forming system, whereas
part of the star-formation activity remains shrouded within more dense and 
dusty 
clouds, as in the Antennae Galaxies, where absorption is ten times higher
when derived from mid-IR than from J, H, K bands, as most massives stars
are not visible at optical wavelengths (\cite{antenne}). This implies that
we can not exclude a SFR as high as ten times what is derived from the optical
 for the cD of A2390.
 Those differences explain why, with a 6.7 $\mu$m/15 $\mu$m ratio of 0.6,
 we can not just consider the cD of A2390 as
an early-type galaxy undergoing simple star-formation in dust lanes as Cen A,
 but most
probably as the place of one or several massive star-bursts which may be
 located in the central globules (see Fig \ref{zoom}).
 Indeed, the study 
of the X-ray image 
of this cluster
demonstrated the presence of one of the strongest
 cooling flows known ($\sim 800$ M$_{\odot}$ yr$^{-1}$ within a cooling radius 
of 200 kpc),
 surrounding the cD galaxy 
(\cite{marguerite}). Giving the size of the mid-IR emitting region in the cD,
 $\sim 20$ kpc, we derive a mass flow of 80 M$_{\odot}$ yr$^{-1}$, 
assuming that 
$\dot{M} \sim r$ (\cite{fabian}).
 Note that 20 kpc is also about the size of the optical filament (Fig. \ref{zoom}). However, our present 
understanding of 
the relationship between mid-IR dust emission and star formation is still too
preliminary to infer quantitative constraints on the IMF or even on the 
heating processes involved in this complex medium.
Finally, the total star-formation rate of $\sim 10$ M$_{\odot}$yr$^{-1}$
 deduced
 from the optical in the filament is clearly a lower limit, and the 
huge quantity of gas needed could be provided by the cooling flow. However, 
despite the fact that 
spiral galaxies are very rare in the core of rich clusters, the hypothesis
of a past merge with a late-type galaxy cannot be formally excluded here, 
which would also
provide gas for some $10^{7-8}$ years.
\subsection{What is new in the arc system of Abell 2390 ?}
After the detection of the giant arc at $z=0.724$ in Abell 370
 (\cite{Metcalfe}), 
observation of the complex arc system of Abell 2390 confirms the
capability of ISOCAM to point up very distant lensed objects.
The giant arc consists of three parts, A at $z=1.033$ (\cite{brenda}), and 
B--C, at $z=0.913$ 
(\cite{Pello}). Near IR imaging already distinguished A from B--C, as A 
was not detected in the K band (\cite{Smail}). HST images revealed that B and 
C are likely two interacting galaxies.
The present ISOCAM images are 
in full agreement with this picture. Although it was not possible to 
estimate properly the 6.7 $\mu$m flux because of blending, the 
15 $\mu$m/6.7 $\mu$m ratio for the B--C 
component is well larger than unity which is indicative of the presence of an 
active star forming region 
in agreement with the strong [OII] line detected in the optical spectrum 
(\cite{Pello}). Except for its lower amplification factor, the case of object D
is very similar.
 Its morphology in the HST images is complex 
with probable signs of interaction and low surface brightness extensions. The 
existence of starburts in the two interacting galaxies is then not a surprise.
The optical and near-IR SEDs of objects B--C and D appear brighter in the
near-IR and fainter in the blue bands compared to A. 
These SEDs can be 
fitted
by different synthetic spectra at $z = 0.913$,
using the GISSEL96 code (Bruzual \&
Charlot, 1998) to approximately constrain the parameters, and 
a single stellar population (instantaneous burst),
 an extinction curve of SMC type (Pr\'evot et al. 1984), and
assuming the Scalo IMF (1986).
 The best fits of the sources B--C and D 
 are obtained with a rest-frame $A_{V} \sim3$ in both cases,
a stable result with respect to metallicity changes. The corrected magnitudes for
objects B--C and D (lensing and absorption) are very similar ($M_B = -20.8$) 
, the total mass involved 
in the burst being $\sim10^{10} M_{\odot}$ in both cases. Despite 
uncertainties on burst age, a 
constant star-forming model gives similar results and
a mean corrected SFR of $40$ to $50\, M_{\odot} h_{50}^{-2} yr^{-1}$.
 According to these results, 
the two lensed sources detected by ISOCAM at $z=0.913$ are strongly reddened star-forming galaxies. In the case of A, there
is no need for a reddening correction to fit the SED.

Finally, the ISOCAM source \#4 detected in both channels 
may be associated with a very faint source in the HST image (I = 23.5), with a 
fuzzy shape. Its 15 $\mu$m/6.7 $\mu$m ratio is very high ($\sim 3$). A photometric redshift 
of $z=0.4^{+0.2}_{-0.08}$ is proposed for this object by techniques described
by Miralles \& Pell\`o (1998).
Even if the results are much more uncertain in this case ($75\%$ confidence),
 the best fit of the SED 
gives $A_{V} \sim$3.5--4.2 in order to explain the high J and K$^\prime$ emission compared to
the optical bands. The corrected SFR is relatively low, $\sim1 M_{\odot} 
h_{50}^{-2} yr^{-1}$. Taking the photometric redshift into
account, the SED of this object, with strong mid-IR emission with respect to its
optical counterpart, is probably dominated by the so-called 
unidentified infrared band emitters,
and its colors are similar to those of the post-starburst companion of M51
 (Boulade et al., 1996).

\subsection{Summary and conclusion}

From deep and high-resolution ISOCAM images of the core of Abell 2390 we
 discovered
active star forming regions in the two most distant lensed galaxies ever 
seen in a cluster by ISO.
This allowed us to support the scenario of two interacting galaxies at 
$z=0.913$ in the
``straight arc'' of A2390, as well as in the other galaxy observed at the 
same redshift. More interesting, we detect
a very faint emission from the cD galaxy at 6.75 $\mu$m,
 compared to other cluster dominant galaxies at similar redshift 
(\cite{next}). But, for the first time, the strong 15 $\mu$m/6.75 
$\mu$m emission ratio flags the presence of a large 
amount of warm dust in the cD, probably 
associated with a very active 
star forming region located within the envelope of the galaxy. Thus,
our observation may further elucidate the fate
of part of accumulating gas in the complex cooling flow radio core 
environment. 

\begin{acknowledgements} 
We are grateful to 
J.-P. Kneib 
for numerous informations on the HST images. Thanks also go to J.-L. Starck,
 H. Aussel and S. Madden for helpful discussions.
The ISOCAM data presented in this paper were analysed using "CIA",
a joint development by the ESA Astrophysics Division and the ISOCAM
Consortium led by the ISOCAM PI, C. Cesarsky, Direction des Sciences de
la Matiere, C.E.A., France.
\end{acknowledgements}

\end{document}